\documentclass[10pt,amsmath,aps,prapplied,twocolumn,superscriptaddress,longbibliography,numbers,sort&compress]{revtex4-2}

\RequirePackage{graphicx}

\RequirePackage{xcolor}

\RequirePackage{siunitx}
\DeclareSIUnit{\dBm}{dBm}

\RequirePackage{nicefrac}

\RequirePackage{booktabs}

\RequirePackage{enumitem}

\RequirePackage[normalem]{ulem}
\RequirePackage{wrapfig}
\RequirePackage{verbatim}

\RequirePackage[colorlinks=true,linkcolor=blue,urlcolor=blue,hyperfootnotes=false,citecolor=black]{hyperref}

\RequirePackage[capitalise]{cleveref}

\newcommand{\DSNR}{\Delta \mathrm{SNR}}
\newcommand{\DSNRBW}{\left\langle \Delta \mathrm{SNR} \right\rangle_{BW}}
\newcommand{\DSNRmean}{\left\langle \Delta \mathrm{SNR} \right\rangle_{\SIrange{4}{8}{\giga\hertz}}}

\newcommand{\GBW}{\left\langle G \right\rangle_{BW}}
\newcommand{\Gmean}{\left\langle G \right\rangle_{\SIrange{4}{8}{\giga\hertz}}}

\newcommand{\Sto}{\mathrm{TWPA} \; S_{21}}
\newcommand{\Stomean}{\left\langle \mathrm{TWPA} \; S_{21} \right\rangle_{\SIrange{4}{8}{\giga\hertz}}}
\newcommand{\Stopl}{S_{21}}
\newcommand{\Stoplmean}{\left\langle S_{21} \right\rangle_{\SIrange{4}{8}{\giga\hertz}}}

\newcommand{\freqrange}{\SIrange{4}{8}{\giga\hertz}}

\newcommand{\BW}{{BW}_{\SI{3}{\decibel}}}

\newcommand{\fp}{f_{\mathrm{p}}}

\newcommand{\Pp}{P_{\mathrm{p}}}

\newcommand{\Bperp}{B_{\mathrm{\perp}}}
\newcommand{\Bparone}{B_{\mathrm{\parallel}, 1}}
\newcommand{\Bpartwo}{B_{\mathrm{\parallel}, 2}}
\newcommand{\Bpar}{B_{\mathrm{\parallel}}}

\newcommand{\Tc}{T_{\mathrm{c}}}

\newcommand{\Bc}{B_{\mathrm{c}}}

\newcommand{\fiftyohm}{\SI{50}{\ohm}}
\newcommand{\threedB}{\SI{3}{\decibel}}

\newcommand{\alphaKI}{\alpha_{\mathrm{KI}}}
\newcommand{\sigman}{\sigma_{\mathrm{n}}}
\newcommand{\sigmaNbTiN}{\sigma_{\mathrm{n}}^{\mathrm{NbTiN}}}
\newcommand{\sigmaNb}{\sigma_{\mathrm{n}}^{\mathrm{Nb}}}
\newcommand{\Deltazero}{\Delta_0}
\newcommand{\gapratio}{2\Delta_0/k_{\mathrm{B}} T_{\mathrm{c}}}

\newcommand{\Teff}{T_{\mathrm{eff}}}
\newcommand{\Tmin}{T_{\mathrm{min}}}
\newcommand{\Tsecond}{T_{\mathrm{2nd}}}

\newcommand{\Zs}{Z_{\mathrm{s}}}
\newcommand{\Rs}{R_{\mathrm{s}}}
\newcommand{\Xs}{X_{\mathrm{s}}}
\newcommand{\Zss}{Z_{\mathrm{s},\mathrm{s}}}
\newcommand{\Zsg}{Z_{\mathrm{s},\mathrm{g}}}
\newcommand{\Rss}{R_{\mathrm{s}}^{(\mathrm{s})}}
\newcommand{\Rsg}{R_{\mathrm{s}}^{(\mathrm{g})}}

\newcommand{\ts}{t_{\mathrm{s}}}
\newcommand{\tg}{t_{\mathrm{g}}}

\newcommand{\Zzero}{Z_{\mathrm{0}}}
\newcommand{\etazero}{\eta_{\mathrm{0}}}

\newcommand{\alphad}{\alpha_{\mathrm{d}}}
\newcommand{\alphac}{\alpha_{\mathrm{c}}}

\newcommand{\Ls}{L_{\mathrm{s}}}
\newcommand{\Ystub}{Y_{\mathrm{stub}}}

\newcommand{\Zenv}{Z_{\mathrm{0}}^{\mathrm{env}}}

\newcommand{\Tsignal}{T_{\mathrm{signal}}}
\newcommand{\Tidler}{T_{\mathrm{idler}}}
\newcommand{\Tp}{T_{\mathrm{p}}}
\newcommand{\omegasignal}{\omega_{\mathrm{signal}}}
\newcommand{\omegaidler}{\omega_{\mathrm{idler}}}

\newcommand{\epsr}{\varepsilon_{\mathrm{r}}}
\newcommand{\epsrsub}{\varepsilon_{\mathrm{r},\mathrm{sub}}}
\newcommand{\epsrsuper}{\varepsilon_{\mathrm{r},\mathrm{super}}}
\newcommand{\tandeltasub}{\tan\delta_{\mathrm{sub}}}
\newcommand{\tandeltasuper}{\tan\delta_{\mathrm{super}}}
\newcommand{\qsub}{q_{\mathrm{sub}}}
\newcommand{\qsuper}{q_{\mathrm{super}}}

\graphicspath{{figures/}}

\begin{document}

\title{Magnetic-Field and Temperature Limits of a Kinetic-Inductance Traveling-Wave Parametric Amplifier}

\author{Lucas~M.~Janssen}
\affiliation{Physics Institute II, University of Cologne, Z\"ulpicher Str. 77, 50937 K\"oln, Germany}
\author{Farzad~Faramarzi}
\affiliation{Jet Propulsion Laboratory, California Institute of Technology, Pasadena, CA 91101, USA}
\author{Henry G.~LeDuc}
\affiliation{Jet Propulsion Laboratory, California Institute of Technology, Pasadena, CA 91101, USA}
\author{Sahil Patel}
\affiliation{Department of Applied Physics and Materials Science, California Institute of Technology, Pasadena, California 91125, USA}
\affiliation{Jet Propulsion Laboratory, California Institute of Technology, Pasadena, CA 91101, USA}
\author{Gianluigi Catelani}
\affiliation{JARA Institute for Quantum Information (PGI-11), Forschungszentrum J\"ulich, 52425 J\"ulich, Germany}
\affiliation{Quantum Research Center, Technology Innovation Institute, Abu Dhabi 9639, UAE}
\author{Peter~K.~Day}
\affiliation{Jet Propulsion Laboratory, California Institute of Technology, Pasadena, CA 91101, USA}
\author{Yoichi~Ando}
\email{ando@ph2.uni-koeln.de}
\affiliation{Physics Institute II, University of Cologne, Z\"ulpicher Str. 77, 50937 K\"oln, Germany}
\author{Christian~Dickel}
\email{dickel@ph2.uni-koeln.de}
\affiliation{Physics Institute II, University of Cologne, Z\"ulpicher Str. 77, 50937 K\"oln, Germany}

\begin{abstract}
  Kinetic-inductance traveling-wave parametric amplifiers (KI-TWPAs) offer broadband near-quantum-limited amplification with high saturation power. 
  Due to the high critical magnetic fields of high-kinetic-inductance materials, KI-TWPAs should be resilient to magnetic fields.
  In this work, we study how magnetic field and temperature affect the performance of a KI-TWPA based on a thin-NbTiN inverse microstrip with a Nb ground plane. 
  This KI-TWPA can provide substantial signal-to-noise ratio improvement ($\DSNR$) up to in-plane magnetic fields of \SI{0.35}{\tesla} and out-of-plane fields of \SI{50}{\milli\tesla}, considerably higher than what has been demonstrated with TWPAs based on Josephson junctions. 
  The field compatibility can be further improved  by incorporating vortex traps and by using materials with higher critical fields. 
  We also find that the gain does not degrade when the temperature is raised to \SI{3}{\kelvin} (limited by the Nb ground plane) while $\DSNR$ decreases with temperature consistently with expectation.
  This demonstrates that KI-TWPAs can be used in experiments that need to be performed at relatively high temperatures.
  The operability of KI-TWPAs in high magnetic fields opens the door to a wide range of applications in spin qubits, spin ensembles, topological qubits, low-power NMR, and the search for axion dark matter.
\end{abstract}

\maketitle

\section{Introduction}

\begin{figure*}[t!]
  \centering  \includegraphics[width=\textwidth]{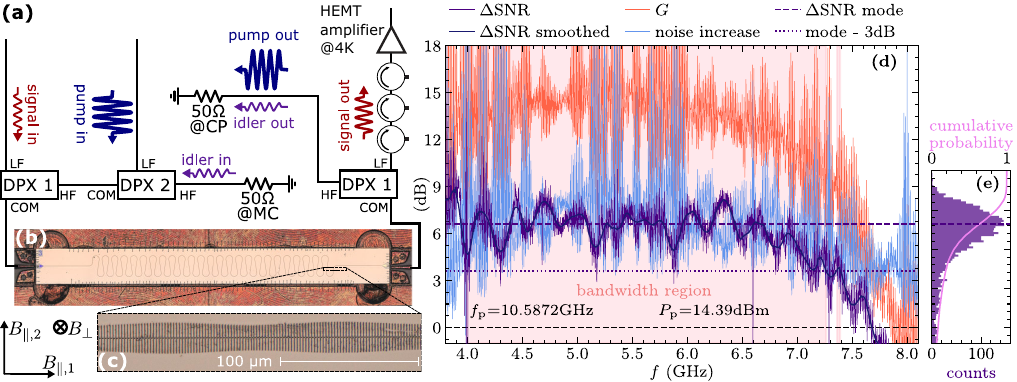}
  \caption{
    \textbf{(a)} Measurement setup with diplexers for coupling the pump tone in and out and idler thermalization.
    After the TWPA, there are circulators and a HEMT amplifier in the cryogenic signal chain.
    \textbf{(b)} Microscope image of the TWPA showing a meandering microstrip. 
    Magnetic field directions $\Bperp$, $\Bparone$, and $\Bpartwo$ are related to the device geometry.
    \textbf{(c)} Zoom-in of microstrip region showing the capacitive loading stubs with modulated length for bandgap engineering.
    \textbf{(d)} Representative SNR improvement $\DSNR$ and gain $G$ curves for the indicated pump settings at zero field and base temperature.
     $\DSNR$ is the difference of gain and added noise. 
     Shaded region indicates \threedB-bandwidth relative to $\DSNR$ mode (with smoothing applied).
    \textbf{(e)} $\DSNR$ histogram and cumulative probability distribution. 
    $\DSNR$ mode corresponds to the maximum slope of the cumulative distribution and is used to define the plateau region.
  }
  \label{fig:figure_1}
\end{figure*}

Kinetic inductance traveling-wave parametric amplifiers (KI-TWPAs)~\cite{ho2012} have emerged as promising technology that can offer large instantaneous bandwidth, high dynamic range, and near-quantum-limited added noise~\cite{Malnou2021,Faramarzi2024,FaramarziTiN2025,Howe2025}. 
They are already becoming established for the readout of arrays of energy-resolving single-photon-counting superconducting detectors, such as Microwave Kinetic Inductance Detectors (MKIDs)~\cite{Bockstiegel2014,Zobrist2019} and Transition Edge Sensors (TES)~\cite{giachero2022}. 
The low added noise makes them in principle suitable for superconducting qubit readout~\cite{CastellanosBeltran2025}, but higher pump powers compared to Josephson TWPAs (J-TWPAs) may present practical challenges.

Although J-TWPAs have undergone substantial development~\cite{Macklin2015, White2015, Planat2020, Renberg2023, Gaydamachenko2025}, KI-TWPAs offer four distinct advantages: higher saturation power, magnetic field resilience, higher-temperature and higher-frequency operation. 
The resilience of KI-TWPAs to magnetic fields can be attributed to two factors: the inherently high critical fields of high-kinetic-inductance materials (e.g., NbTiN or TiN as opposed to predominantly aluminum-based Josephson junctions) and the absence of the Fraunhofer effect that in Josephson junctions can cause oscillatory suppression of the critical current at millitesla-scale fields~\cite{Janssen2024}. 
SQUID-based J-TWPAs~\cite{Planat2020, Gaydamachenko2025} would often be sensitive to sub-millitesla out-of-plane fields. 
The impressive field resilience of kinetic inductance materials has already been shown in recent works on cavity-based kinetic inductance parametric amplifiers ~\cite{Khalifa2023,Xu2023, Splitthoff2024, Zapata2024, Frasca2024, Vaartjes2024}, however, these amplifiers suffer from lower bandwidths and saturation power compared to the KI-TWPAs due to their resonant nature. 

There is already a growing number of experiments that employ TWPAs and require relatively high magnetic fields at the samples~\cite{Uilhoorn2021, Wang2023, Bartram2023, DiVora2023, Elhomsy2023,Aghaee2025nature,aghaee2025}. In addition, there are many examples of experiments that would benefit from a TWPA, e.g. in superconducting qubits~\cite{Luthi18,Schneider2019, Pita-Vidal2020, Kringhoj21, Krause2022,Krause2024, Gunzler2025}, quantum dots and spin qubits~\cite{zheng2019, Schaal2020, deJong2021, Elhomsy2023}, spin ensembles or low-power NMR~\cite{Bienfait2016, Wang2023, Vine2023}, topological qubits~\cite{Aghaee2025nature,aghaee2025}, as well as hybrid setups including mechanical~\cite{Kounalakis2020, Bera2021} and magnonic degrees of freedom~\cite{Tabuchi2015, Kounalakis2022}, and the search for axion dark matter~\cite{chaudhuri2024introducing,Bartram2023, DiVora2023,Ramanathan2023}.

KI-TWPAs based on NbTiN can also operate at higher temperatures because of the high critical temperature of NbTiN, making them more versatile. 
In the future, they might replace commonly used cryogenic HEMT amplifiers in some use cases, with similar performance for a fraction of the power dissipation~\cite{Malnou2022}.

Here, we characterize the performance of a KI-TWPA, with NbTiN as the microstrip layer and Nb as a ground layer, as a function of temperature and different magnetic field directions.
We measure both gain and signal-to-noise (SNR) improvement $\DSNR$, which quantify improvement over the subsequent amplification chain.
We find that the KI-TWPA maintains a significant improvement in SNR ($>$ 3 dB) up to \SI{1.2}{\kelvin}, although a higher temperature increases the idler noise because the input port of the device is terminated at the variable temperature stage. 
The temperature dependence of the unpumped transmission of the TWPA can be modeled using Mattis-Bardeen theory with temperature-dependent suppression of the superconducting gap.
Similarly, $\DSNR$ can be modeled assuming an ideal four-wave-mixing parametric amplifier cascaded with a second amplifier, demonstrating near-quantum-limited noise.\\

We next study the effect of magnetic fields, both in-plane and perpendicular.
We report a one order of magnitude improvement in field resilience over J-TWPAs in the bottleneck field directions.
Critically, $\DSNR$ measurements reveal that practical amplifier performance can degrade before gain degradation becomes apparent, emphasizing the importance of measuring both metrics.
The magnetic field dependence shows behavior that cannot be explained simply in terms of gap suppression by the field. 
In fact, considerable hysteresis with the magnetic field indicates vortex dynamics in the niobium ground plane but likely also in the narrow center conductor play a dominant role.
While we do not develop a quantitative model that includes vortex dynamics, the measured performance thresholds provide clear operational guidelines.
With the demonstrated field resilience, KI-TWPAs can be incorporated into most experimental setups where the amplifier can be mounted at some distance from the magnet~\cite{Janssen2024}, especially when stray fields are predominantly in-plane.

\section{Experimental details}

An optical image of the device is shown in \cref{fig:figure_1}(b), with a zoom-in showing the inverse microstrip with modulated capacitive stubs (\cref{fig:figure_1}(c)). The device design is identical to the one presented in Ref.~\cite{Faramarzi2024}.
Magnetic field directions relative to the TWPA chip are also indicated in~\cref{fig:figure_1}.
In the following, we focus on the out-of-plane direction, $\Bperp$, and the in-plane direction along the long axis of the chip, $\Bparone$.

Using a combination of diplexers (DPX1 \& DPX2) before the TWPA, we supply the pump tone and signal tones to the device. Additionally, we provide a cold termination at the input of the device over the idler band to thermalize it to the lowest possible temperature. Using a third diplexer (also labeled DPX1, since it is identical) at the output of the device, we terminate the idler frequencies and the pump tone at a higher dilution refrigerator stage to reduce the heat load on the mixing chamber. A detailed description of the device and measurement setup is provided in \cref{sec:device_description_and_setup}.

In the following, we need figures of merit for the amplifier performance to characterize the temperature and field dependence.
We will focus on the insertion loss, the true gain, the SNR improvement, and the bandwidth.
The way we define and calculate these figures of merit from the data is illustrated in \cref{fig:figure_1}(d,e).
Calibration measurements with a short cable in place of the TWPA provide a reference for calculating the true gain and added noise from the measured $\Stopl$ and noise spectrum.
True gain is the net signal amplification relative to this bypass connection, inherently accounting for the device's insertion loss.
We calculate $\DSNR$ as the difference between gain and added noise.
Crucially, we optimize the pump power $\Pp$ and frequency $\fp$ by maximizing the mean $\DSNR$ in the \freqrange~band,  $\DSNRmean$ (for data on the pump optimization see \cref{sec:twpa_pump_settings_and_optimization}).
For the field and temperature dependence, we optimize the pump settings at every temperature or magnetic field.
We then extract the typical value of $\DSNR(f)$ by approximating the $\DSNR$ mode —the most commonly occurring value.
\cref{fig:figure_1}(e) shows a measured data histogram and cumulative probability distribution.
Relative to the $\DSNR$ mode, we define the \threedB-bandwidth $\BW$ by smoothing the $\DSNR$ data and finding ranges of frequencies where this signal is above a \threedB-threshold.
The bandwidth regions need not be continuous; when referring to $\BW$ in frequency units, we sum up disjoint frequency regions.
However, the bandwidth is typically continuous for the optimized pump settings.
All quoted bandwidths are based on $\DSNR$, not the gain profiles.
We then show the mean of $\DSNR$ and true gain within the bandwidth: $\DSNRBW$ and $\GBW$.
The error bars in the plots below represent the 25th and 75th percentiles of the data within the bandwidth.

\section{Temperature dependence}

\begin{figure}
  \centering
  \includegraphics[width=\columnwidth]{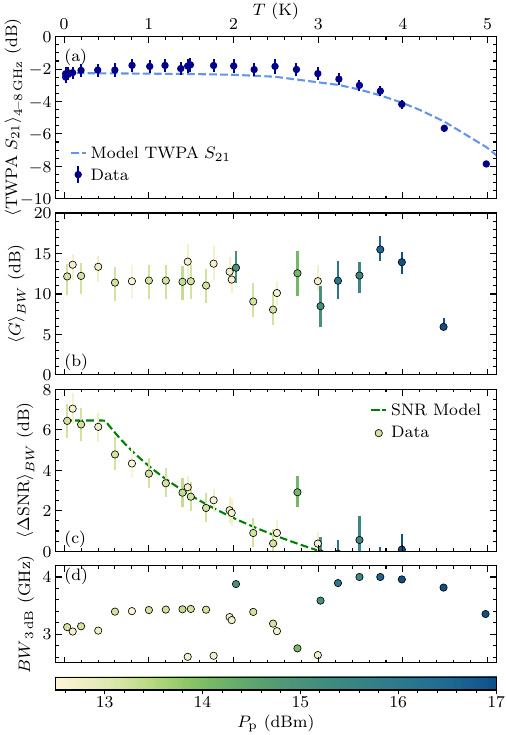}
  \caption{
    \textbf{(a)} $\Stomean$ versus temperature $T$ showing plateau up to about \SI{3}{\kelvin}.
    \textbf{(b)} $\DSNRBW$ versus $T$ showing a decrease in line with a model based on an ideal four-wave-mixing parametric amplifier with input noise. 
    Above $T=\SI{2.5}{\kelvin}$, it drops below \SI{0}{\decibel}.
    \textbf{(c)} True gain within bandwidth $\GBW$ versus $T$ showing a plateau up to \SI{4}{\kelvin}. 
    Around \SI{3}{\kelvin}, the $\Pp$ (optimized at every temperature) increases, likely compensating for the increased insertion loss.
    \textbf{(d)} $\BW$ versus $T$. Bandwidth increases when optimized $\Pp$ increases at high $T$ and then drops.
  }
  \label{fig:twpa_SNR_vs_T}
\end{figure}

We begin by discussing the KI-TWPA temperature dependence and the
model that describes $\Sto$ as a function of $T$; the model provides insight into the underlying physics and will help better understand the magnetic field dependence. 
\cref{fig:twpa_SNR_vs_T}\textbf{(a)} shows $\Stomean$, the transmission calibrated against a bypass cable, as a function of temperature $T$. 
There is no significant change in the transmission of the device up to \SI{3}{\kelvin}. 
Here we show only the $\Stomean$ data for a reduced temperature range, where the TWPA still provides gain; the full dependence on frequency and temperature can be found in \cref{sec:twpa_insertion_loss_vs_temperature}.

We model the temperature dependence using Mattis-Bardeen theory of the AC conductivity~\cite{MattisBardeen1958} taking into account the temperature dependence of the gap (\cref{sec:twpa_model}) and find qualitative agreement.
The surface impedance of both superconductors (Nb and NbTiN) is calculated using this model, which reveals that changes in ground plane surface resistance as a function of temperature dominate the change in $\Stomean$.
Thus, a fully NbTiN-based device would likely operate beyond \SI{4}{\kelvin} with minimal loss.

Now we turn from the insertion loss to the TWPA performance based on the aforementioned figures of merit.
\cref{fig:twpa_SNR_vs_T}\textbf{(b)} and \textbf{(c)} show $\GBW$ and $\DSNRBW$ versus $T$.
The gain maintains similar values up to \SI{4}{\kelvin}, but the optimized $\Pp$ starts to increase around \SI{3}{\kelvin}, consistent with the increasing insertion loss.
The bandwidth initially shows a slight increase, and then increases when optimal $\Pp$ increases at high $T$.

We model the temperature-dependent $\DSNR$ of the parametric amplifier cascaded with a second amplifier.
In this case the second amplifier would be the HEMT shown in \cref{fig:figure_1}; we expect the subsequent amplifiers add minimal noise due to the gain of the TWPA and the HEMT amplifier. 
The parametric amplifier provides a true gain $G$ with noise from the signal and idler ports at frequencies given by $2\fp = f_{\text{signal}} + f_{\text{idler}}$. 
The noise temperature contributions at the signal and idler tones are given by $\Tsignal = \frac{\hbar\omegasignal}{k_B}(1 + 2n(\omegasignal, \Teff))$ and $\Tidler = \frac{\hbar\omegaidler}{k_B}(1 + 2n(\omegaidler, \Teff))$, where $n(\omega, \Teff)$ is the Bose-Einstein occupation number and $\Teff = \max(T, \Tmin)$ includes a minimum-noise floor $\Tmin$ to account for
the low-temperature plateau in $\DSNR$ (see \cref{fig:twpa_SNR_vs_T}\textbf{(b)}).
Using Friis' formula~\cite{Friis1946}, the total parametric amplifier noise temperature $\Tp = \Tsignal + \Tidler$ gives an SNR improvement
\begin{equation*}
\Delta\text{SNR}_{\text{dB}} = 10\log_{10}\left(\frac{1}{\Tp/\Tsecond + 1/G_\mathrm{linear}}\right).
\end{equation*} 
Using the above model and assuming a fixed TWPA gain of \SI{12}{\decibel} (matching measured values) results in a good agreement with data.
It yields a second amplifier noise temperature of \SI{13}{\kelvin} and $\Tmin = \SI{0.48}{\kelvin}$. 
The \SI{13}{\kelvin} estimate significantly exceeds the nominal specified noise temperature of \SI{3}{\kelvin} for the HEMT, suggesting attenuation between the amplifiers is not negligible. 
Using the Friis formula for lossy cascaded systems, approximately \SI{3.9}{\decibel} of loss would increase the effective noise temperature from \SI{3}{\kelvin} to \SI{13}{\kelvin}.
The specified total insertion loss of the diplexer plus circulators would be around \SI{1.6}{\decibel} and we have 8 SMA connectors between the sample and HEMT which could account for another \SI{1.6}{\decibel}, so the loss is on the high end of what we expect, but not implausible, especially given that the HEMT could also be slightly sub-optimally biased.

\section{In-plane magnetic field dependence}
\label{sec:in-plane_magnetic_field_dependence}

\begin{figure}
  \centering
  \includegraphics[width=\columnwidth]{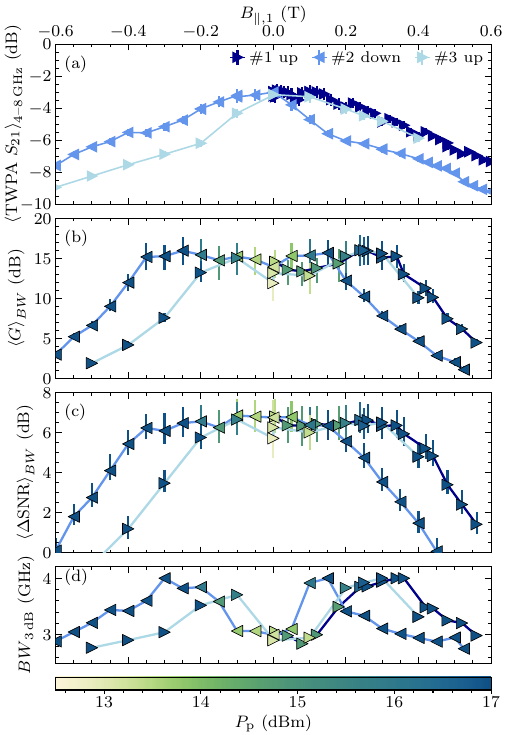}
  \caption{
    \textbf{(a)} $\Bparone$ dependence of $\Sto$. 
    The sweep direction is color-coded to reveal the hysteresis.
    \textbf{(b)} Corresponding gain curves at different $\Bparone$. 
    Pump settings are optimized at every field; data is selected for optimum $\DSNRmean$. 
    Data points are color-coded for $\Pp$ and lines indicate sweep direction: higher $\Bparone$ requires higher $\Pp$, consistent with increased insertion loss.
    The hysteresis is also visible for the gain curves.
    \textbf{(c)} $\DSNRBW$ versus $\Bparone$, following gain curves.
    \textbf{(d)} $\BW$ at different $\Bparone$ showing initial increase then decrease. $\BW$ maximized at $\Bparone\approx\SI{0.2}{\tesla}$ while gain and $\DSNRBW$ similar to zero field values. $\BW$ shows clear hysteresis.
  }
  \label{fig:twpa_SNR_vs_Bpar1}
\end{figure}

The in-plane magnetic field dependence of the TWPA is shown in \cref{fig:twpa_SNR_vs_Bpar1}.
\cref{fig:twpa_SNR_vs_Bpar1}\textbf{(a)} shows $\Stomean$ versus $\Bparone$, color-coded by sweep direction to reveal hysteresis.
For in-plane fields, device insertion loss recovers its initial value when scanning back to zero field.
Wider range scans and detailed insertion loss data are presented in \cref{sec:twpa_insertion_loss_vs_magnetic_field}, showing that $\Sto$ is fully suppressed around \SI{1.5}{\tesla}, close to the critical field $\Bc$ of niobium (cf. \cref{fig:dc_measurements_Tc_Bc}).
Data for the $\Bpartwo$ direction shows less pronounced suppression and hysteresis, indicating possible magnet-to-sample misalignment as it is not obvious that the two directions should be different based on the sample geometry.

The Mattis-Bardeen model that describes temperature dependence can be modified to include field-dependent gap suppression; it does not adequately describe in-plane field dependence of $\Sto$: simple gap suppression with field would only predict an abrupt change near $\Bc$, while we observe gradual degradation at much lower fields.
Still, the model qualitatively reveals some of the effects that the field has on the device.
In the model, the surface inductance changes by roughly an order of magnitude, both with temperature and magnetic field, but inductance changes do not dominate the $\Sto$ suppression.
Moreover, unlike for the temperature, the magnetic field-induced gap suppression alone does not increase surface resistance (except near $\Bc$). 

Two effects can explain the additional change in surface resistance we see in experiment: the presence of vortices and their motion in the niobium ground plane (partly due to field misalignment), or screening currents from flux penetration that enhance pair breaking.
Advanced models that account for vortices can describe field dependence in superconducting resonators~\cite{Kwon2018}, and field-dependent gap changes could help constrain parameters as in Ref.~\cite{Janssen2024}. 
The reasonable temperature agreement suggests the modeling approach is sound but requires better niobium ground plane impedance modeling.
However, unknown misalignment complicates this since $\Bperp$ and $\Bpar$ dependencies differ significantly.
We therefore focus on measured performance thresholds rather than pursuing quantitative modeling of the magnetic field dependence.

\cref{fig:twpa_SNR_vs_Bpar1} \textbf{(b)} shows the corresponding gain curves at different $\Bparone$ values.
The gain is optimized at every field value and the plotted data is selected for optimum $\DSNRmean$.
Data points are color-coded for $\Pp$, revealing that, in general, at higher $\Bparone$, higher $\Pp$ is required, in line with the increased insertion loss.
\cref{fig:twpa_SNR_vs_Bpar1} \textbf{(c)} shows $\DSNRBW$ as a function of $\Bparone$, which follows the gain curves.
\cref{fig:twpa_SNR_vs_Bpar1} \textbf{(d)} shows $\BW$ at different $\Bparone$ values, showing an initial increase with field followed by a decrease.
At $\Bparone \approx \SI{0.3}{\tesla}$, $\BW$ is maximized, while the gain and $\DSNRBW$ remain similar to their zero-field values.
Increased bandwidth likely results from field-induced losses suppressing internal reflections and impedance mismatches within the TWPA.
Given that the net gain $\GBW$ and $\DSNRBW$ remain similar and the bandwidth increases, while $\Stomean$ slightly decreases these losses are obviously compensated.
The $\BW$ shows clear hysteresis correlated with changed insertion loss (likely correlating with the larger optimum $\Pp$).

\section{Out-of-plane magnetic field dependence}
\label{sec:out-of-plane_magnetic_field_dependence}

\begin{figure}
  \centering
  \includegraphics[width=\columnwidth]{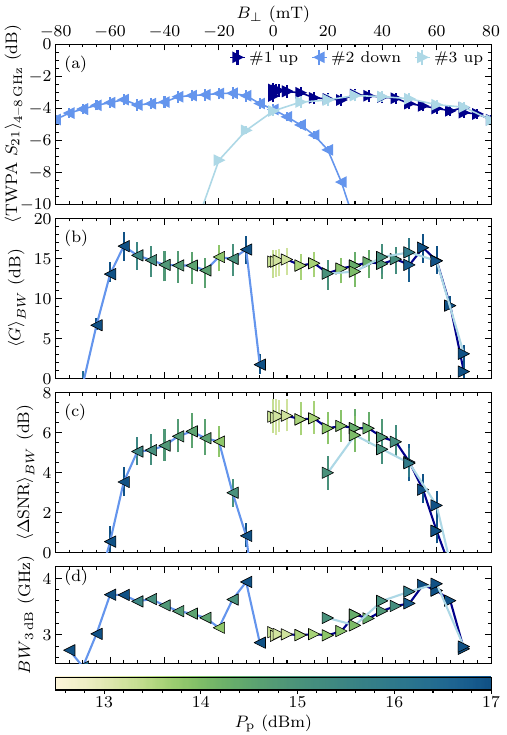}
  \caption{
    \textbf{(a)} $\Bperp$ dependence of $\Stomean$. 
    Hysteresis much more pronounced than for $\Bparone$.
    \textbf{(b)} Corresponding gain curves at different $\Bperp$. 
    Gain remains stable until a steep drop around \SI{60}{\milli\tesla}. Upon downward sweep, gain returns only past going through zero field around $\Bperp \approx \SI{-10}{\milli\tesla}$ and behaves similarly for the subsequent upward sweep.
    \textbf{(c)} $\DSNR$ versus $\Bperp$ already decreases while gain remains stable, suggesting added noise consistent with vortices effectively raising the amplifier temperature. 
    Initial $\DSNR$ is not recovered in subsequent sweeps.
    \textbf{(d)} $\BW$ at different $\Bperp$ increasing until a sudden downturn at \SI{60}{\milli\tesla}. 
    Hysteresis is not pronounced for $\BW$.
  }
  \label{fig:twpa_SNR_vs_Bperp}
\end{figure}

The out-of-plane magnetic field dependence of the TWPA is shown in \cref{fig:twpa_SNR_vs_Bperp}.
\cref{fig:twpa_SNR_vs_Bperp} \textbf{(a)} shows the $\Bperp$ dependence of $\Stomean$.
Hysteresis is much more pronounced than for $\Bparone$.
The $\Sto$ for a wider field range can be found in \cref{sec:twpa_insertion_loss_vs_magnetic_field}.
Overall, this hysteresis is similar to what has been observed in superconducting resonators~\cite{Bothner2012}.
It seems that after scanning to high $\Bperp$, the initial $\Sto$ is not fully recovered but transmission does come back when scanning slightly past zero field.
We have DC measurements of $\Bc$ along the $\Bperp$ axis for a film similar to the niobium ground plane (\cref{sec:dc_characterization_of_material_tc_and_bc}); we estimate $\Bc$ in the range \SIrange{1.5}{1.9}{\tesla}, much larger than fields where transmission is suppressed here.
This, combined with the pronounced hysteresis, indicates the dominant physics is due to vortices rather than gap suppression.
While theoretical frameworks exist for modeling field-dependent superconductor losses and hysteresis~\cite{Bothner2012} and there are field-dependent complex surface impedance descriptions that include vortices (see Ref.~\cite{Kwon2018}, whose approach could be integrated into ours in \cref{sec:twpa_model}), these models require many parameters (vortex viscosity, pinning potentials) and assumptions, e.g. about the current distribution in the ground plane.
Observed catastrophic flux avalanches~\cite{Nulens2023} could also complicate quantitative modeling, which we therefore do not pursue here.

\cref{fig:twpa_SNR_vs_Bperp} \textbf{(b)} shows the gain curves at different $\Bperp$.
The gain remains stable up to around \SI{60}{\milli\tesla}, where a steep drop is observed.
This drop could be due to vortices entering the NbTiN microstrip: the field $H_s$ at which vortices are stable in the center of a strip of width $W$ and coherence length $\xi$ is given by (see \cite{Song2009} and references therein)
\begin{equation}
    H_s = \frac{2\Phi_0}{\pi W^2}\ln \frac{2W}{\pi \xi}
\end{equation}
which in our sample ($W=340\,$nm, $\xi=4.9\,$nm based on the critical field value extrapolated at low temperature, see Appendix~\ref{sec:dc_characterization_of_material_tc_and_bc}) is of order 43~mT. If cooling in a field below this value, as done in our experiment, no vortices are expected to be present in the strip; an energy barrier impeding vortex entry is present even above this field, but it decreases with increasing field or in the presence of currents, which in our case are due to the pump. Therefore, at sufficiently high field $H \gtrsim H_s$ a configuration with vortices in the microstrip could be realized.
Upon the downward sweep, gain only returns around $\Bperp \approx \SI{-10}{\milli\tesla}$, similarly for the subsequent upward sweep.
\cref{fig:twpa_SNR_vs_Bperp} \textbf{(c)} shows SNR improvement as a function of $\Bperp$ which already decreases while the gain remains stable, suggesting the vortices raise the effective temperature of the TWPA.
The $\DSNR$ starts to steeply drop already around \SI{50}{\milli\tesla}.
The initial SNR improvement is not recovered in the subsequent sweeps.
\cref{fig:twpa_SNR_vs_Bperp} \textbf{(d)} shows bandwidth at different $\Bperp$ increasing until a sudden downturn at \SI{60}{\milli\tesla}.
Hysteresis is not pronounced for the bandwidth, likely because the bandwidth is not directly field dependent but rather depends on the optimized $\Pp$.

For the $\Bparone$ dependence, the $\Gmean$ and $\DSNRmean$ plateaus align, while for $\Bperp$, $\DSNR$ decays before the gain does.
This indicates vortices may effectively raise device temperature rather than just adding attenuation.
This is also apparent when comparing the $\Stomean$ values where $\GBW$ plunges for in-plane versus out-of-plane fields. 

\section{Conclusion and outlook}

We have shown that a kinetic inductance TWPA provides substantial SNR improvement in in-plane magnetic fields of up to \SI{0.35}{\tesla} and in out-of-plane fields of up to \SI{50}{\milli\tesla}, an order-of-magnitude improvement over the bottleneck field shown for J-TWPAs~\cite{Janssen2024}.
With this operating range, it could be used in high-field experiments without magnetic shielding if positioned at some distance from the magnet center.
The bandwidth can actually improve at intermediate in-plane fields with similar gain and added noise.
We attribute this to increased TWPA loss suppressing pump reflections and improving device performance.

While this KI-TWPA is not as field-compatible as some demonstrated parametric amplifiers~\cite{Khalifa2023,Xu2023,Frasca2024,Splitthoff2024,Zapata2024}, it offers larger bandwidth and saturation power.
The field compatibility is mostly limited by vortex losses due to the absence of vortex trapping structures~\cite{Bothner2011, Borsoi2019}.
Using an NbTiN or NbN ground plane with vortex traps could improve field compatibility to several tesla in-plane.
Out-of-plane fields will remain limiting: even thin-film NbTiN resonators with \SI{100}{\nano\meter} width and distant ground planes show order-of-magnitude quality factor decreases at \SI{0.4}{\tesla}~\cite{Samkharadze2016}. 

The temperature dependence of the KI-TWPA shows effective operation up to \SI{3}{\kelvin}, consistent with expectations given that signal and idler input noise also scale with temperature.
SNR improvement over the HEMTs is lost around \SI{3}{\kelvin} for this reason.
Mounting at the still stage (\SI{\sim 1}{\kelvin}) would still provide significant improvement over a HEMT, which has about 50 times the power consumption. 
Moreover, the TWPA pump power does not necessarily have to be completely dissipated at the same temperature stage, but can be terminated elsewhere as in this work.
Similarly, the idler and signal band do not have to be thermalized at the same stage as the TWPA.
This flexibility, combined with high gain, saturation power and bandwidth, positions KI-TWPAs as promising technology for various quantum and classical applications.
The demonstrated magnetic field and temperature performance highlights KI-TWPA potential to replace or complement existing amplifiers in demanding experimental setups.

\begin{acknowledgments}
We thank Timur Zent for help with assembly of the cryogenic setup. 
This project has received funding from the European Research Council (ERC) under the European Union’s Horizon 2020 research and innovation program (grant agreement No 741121) and was also funded by the Deutsche Forschungsgemeinschaft (DFG, German Research Foundation) under CRC 1238 - 277146847 (Subproject B01) as well as under Germany’s Excellence Strategy - Cluster of Excellence Matter and Light for Quantum Computing (ML4Q) EXC 2004/1 - 390534769. Devices were fabricated at the Jet Propulsion Laboratory, California Institute of Technology,  under a contract with the National Aeronautics and Space Administration (80NM0018D0004).
\end{acknowledgments}

\section*{author contributions}

The project was conceived by C.D. and Y.A. in coordination with F.F. and P.D..
The TWPAs were designed and fabricated by P.D., F.F., and H.G.L. at JPL.
The measurements were performed by L.M.J. and C.D. in Cologne with support from F.F. and P.D.. The DC magnetic measurements of the superconducting films were performed by S.P at Caltech.  
The data was analyzed by L.M.J. and C.D. with support from F.F. and P.D..
G.C. provided theory support for the project.
The manuscript was written by L.M.J., F.F. and C.D. with input from all coauthors.

\section*{Software}

The setup was controlled based on \href{https://github.com/QCoDeS/Qcodes}{QCoDeS} drivers and logging~\cite{Qcodes}, while the measurements were run using \href{https://gitlab.com/quantify-os/quantify-core}{Quantify-core}~\cite{rol2021quantify}.

\section*{Data availability}

Datasets and analysis as well as code for the TWPA model are made available as python files and jupyter notebooks that create the figures of this manuscript on Zenodo with DOI \href{https://doi.org/10.5281/zenodo.17148287}{10.5281/zenodo.17148287}.

\appendix

\section{Device description and measurement setup}
\label{sec:device_description_and_setup}

The device is a kinetic inductance traveling wave parametric amplifier (KI-TWPA) fabricated at the NASA Jet Propulsion Laboratory, similar to the device described in Ref.~\cite{Faramarzi2024}, based on design ideas outlined in Ref.~\cite{Shu2021}.
It consists of a microstrip transmission line with an NbTiN center conductor (\SI{35}{\nano\meter} thick, \SI{340}{\nano\meter} wide) and an Nb ground plane (\SI{350}{\nano\meter} thick).
Neither the center conductor (which has wider areas at the launchers) nor the ground plane incorporates vortex trapping structures.
The kinetic inductance at low field and temperature predominantly stems from the center conductor.

The device is mounted in a copper box that is thermally anchored to the mixing chamber of an Oxford Instruments Triton 400 dilution refrigerator inside a fast-loading puck.
A \SI{6}{\tesla}-\SI{1}{\tesla}-\SI{1}{\tesla} vector magnet provides magnetic fields, with the \SI{6}{\tesla} axis corresponding to $\Bparone$ (see \cref{fig:figure_1}). 
The alignment of the device to the magnetic field axes is imperfect; based on previous experiments, misalignment is estimated below \SI{2}{\degree}~\cite{Krause2022, Krause2024}, but could be greater here because the TWPA might not be perfectly aligned to the box, as it is glued in.
Some observed hysteresis is likely intrinsic to the vector magnet but most of the observed effects originate from the device itself.

\begin{figure}
  \centering
  \includegraphics[width=\columnwidth]{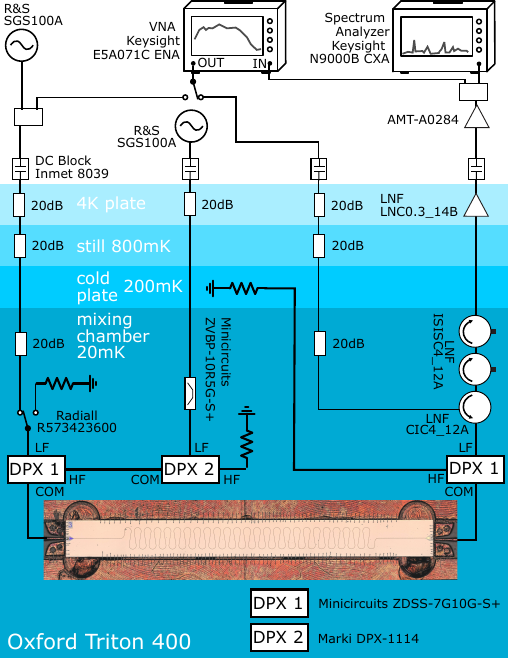}
  \caption{
    Schematic of the cryogenic measurement setup, including input/output lines, attenuators, circulators, and amplification chain.
    The circulator at the TWPA output enables transmission and reflection measurements without internal switching.
    Signal, pump and idler routing through diplexers is necessary due to the high pump power and in order to thermalize the idler input to the mixing chamber.
    Diplexers marked DPX1: low band up to \SI{7.5}{\giga\hertz}, high band \SIrange{10.5}{20}{\giga\hertz}. DPX2: low band up to \SI{11}{\giga\hertz}, high band \SIrange{14}{30}{\giga\hertz}.
  }
  \label{fig:wiring_diagram}
\end{figure}

The measurement setup is shown in \cref{fig:wiring_diagram}.
Two diplexers (DPX1) sandwich the TWPA to couple the pump in and out, preventing saturation of subsequent amplifiers.
These commercial diplexers are not specified for cryogenic operation, but do not show strong changes in transmission during cooldowns.
An additional diplexer (DPX2) thermalizes the idler band at the TWPA input to the mixing chamber.
The frequency ranges of the diplexers limit $\fp$ to approximately \SIrange{9}{11}{\giga\hertz}.

The pump power $\Pp$ is specified throughout this work as the power at the microwave source (approximately \SI{+15}{\dBm}).
The line has one attenuator of \SI{20}{\decibel} at the \SI{4}{\kelvin} stage, but including approximately \SI{20}{\decibel} of losses in stainless steel coaxial cable and additional components, $\Pp$ at the TWPA input is approximately \SI{-25}{\dBm}.
We did not prioritize calibrating lines for accurate device power estimation.

The thermalization of the termination of the pump tone after the TWPA requires careful consideration for KI-TWPAs.
The significant pump power cannot be thermalized at the mixing chamber as it would exceed cooling capacity even in large cryostats.
We attached thermalization to the cold plate, though in retrospect, still-stage thermalization would be preferable to reduce the cold plate heat load.
As the mixing chamber temperature also increases with pump power, we capped $\Pp$ at \SI{17}{\dBm} to maintain base temperatures below \SI{50}{\milli\kelvin}.

For calibration, we measured a ``background transmission'' through our setup with a short cable replacing the TWPA in the loading puck.
This background was subtracted to calculate the net TWPA gain and the added noise.
To exclude that temperature or field affects other components of the setup, we performed temperature and field scans in this bypass configuration (see \cref{sec:field_and_temperature_dependence_of_transmission_without_twpa}).
We found negligible temperature and magnetic field dependence of transmission in the \freqrange\,band. We attribute the small changes in transmission to the circulators.

\section{TWPA pump settings and optimization}
\label{sec:twpa_pump_settings_and_optimization}

\begin{figure*}
  \centering
  \includegraphics[width=2\columnwidth]{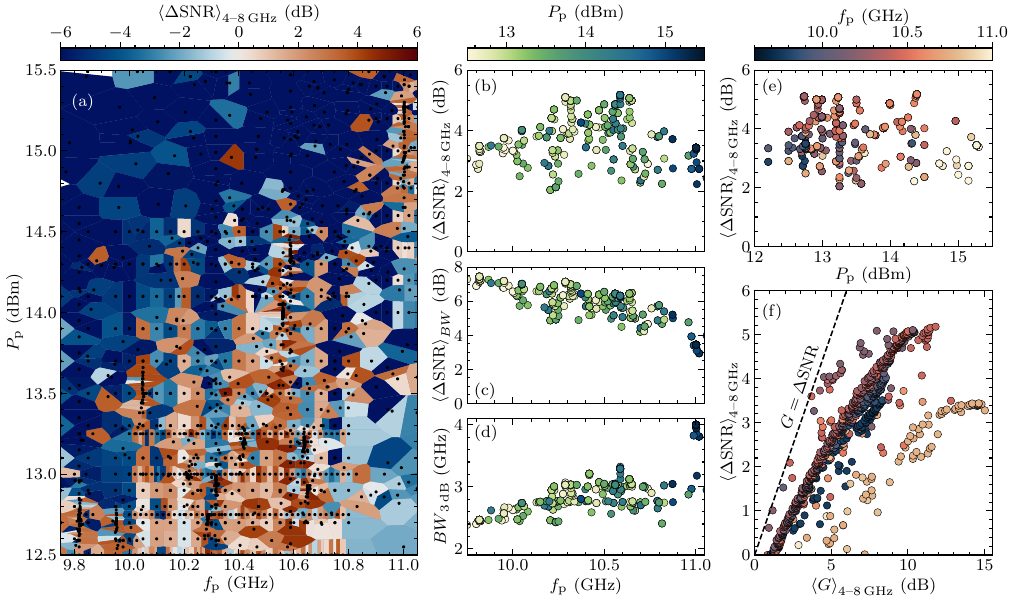}
  \caption{
  \textbf{(a)} Mean SNR improvement in the \freqrange band $\DSNRmean$ versus pump power $\Pp$ and frequency $\fp$ at zero field.
  Measurements are indicated with black points.
  The color map indicates the SNR improvement of the nearest measured point, revealing a chaotic landscape.
  \textbf{(b)-(d)} show $\DSNRmean$, the mean within the bandwidth $\DSNRBW$ and the bandwidth $\BW$, for the same data points as a function of $\fp$ with $\Pp$ color-coded.
  With increasing $\fp$, $\BW$ can be improved, but the $\DSNRBW$ is reduced.
  This leads to some optimum frequencies for $\DSNRmean$.
  \textbf{(e)} $\DSNRmean$ as a function of $\Pp$ with $\fp$ color-coded shows that there is no clear optimum power range.
  \textbf{(f)} SNR improvement vs gain showing clearly that for the bulk of the data there is a clear relationship between the two consistent with a relatively fixed TWPA noise power.
  The higher $\fp$ data shows a different trend, suggesting that the added noise for higher $\fp$ and $\Pp$ is stronger.
  }
  \label{fig:twpa_optimization_summary}
\end{figure*}

To optimize TWPA performance, we measured $\DSNRmean$ for various $\Pp$ and $\fp$ settings at base temperature and zero field.
The results are summarized in \cref{fig:twpa_optimization_summary}.
The choice of $\DSNRmean$ as the figure of merit is practical: the lower limit (\SI{4}{\giga\hertz}) is imposed by the circulator specifications, while this setup achieves poor SNR improvement above \SI{8}{\giga\hertz}.
This figure of merit is a good compromise because it rewards both a larger bandwidth and a better overall SNR improvement.
The data show that the SNR improvement exhibits a chaotic landscape rather than a simple $\Pp$ and $\fp$ dependence, which complicates the optimization of these settings.

We used the Nelder-Mead algorithm~\cite{NelderMead} to find optimal $\Pp$ and $\fp$ for every magnetic field and temperature, using $\DSNRmean$ as the figure of merit. 
$\Pp$ was limited to a maximum of \SI{17}{\dBm} to prevent excessive heating of the fridge. 
We used multiple starting points during the field and temperature scans to avoid being stuck in local maxima. 
The results of different optimizations from slightly varied initial conditions were consistent, showing that we actually find robust maxima. 

Occasionally $\DSNRmean$ exceeded $\Gmean$, occurring when the device switches state between the gain and noise measurements under strong drive.
Thus, there are occasional outliers in the $\DSNRmean$ datasets, but we generally repeated measurements and the switches are rare and not systematic.
\cref{fig:twpa_optimization_summary}(f) shows the general relationship between mean gain and $\DSNRmean$ characteristic of the TWPA, with some outliers.
High pump frequency data follow a curve with lower $\DSNRmean$ saturation, possibly due to heating effects at higher required pump powers.

\section{TWPA insertion loss vs temperature}
\label{sec:twpa_insertion_loss_vs_temperature}

\begin{figure*}
  \centering
  \includegraphics[width=\textwidth]{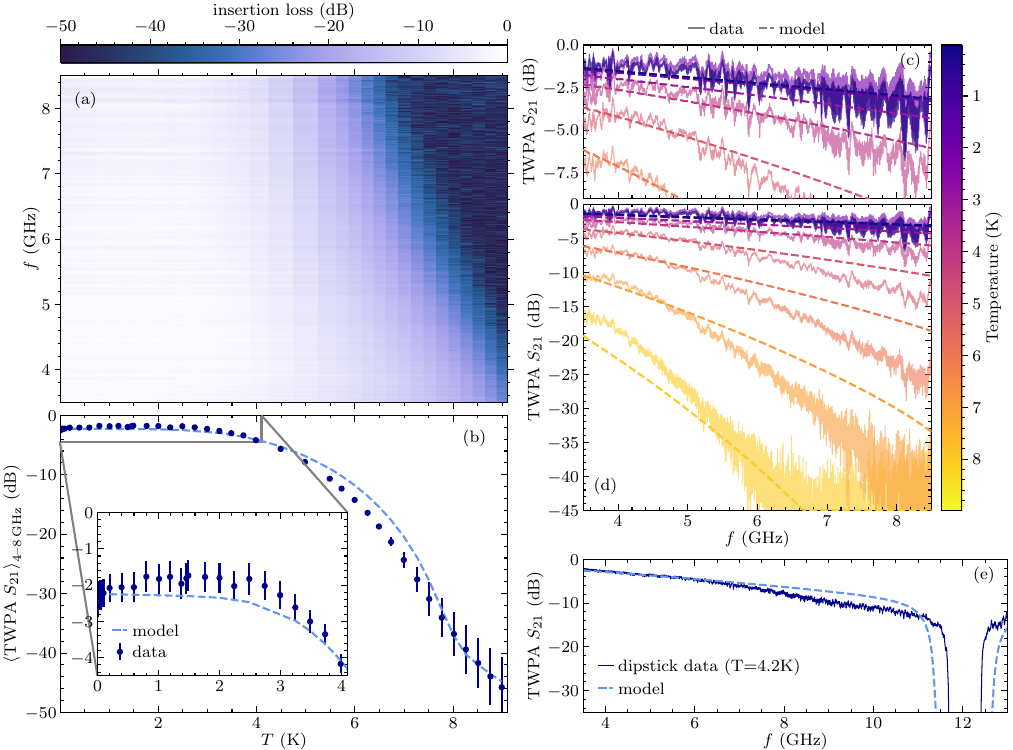}
  \caption{
    \textbf{(a)} $\Sto$ as a function of temperature $T$ and $f$.
    \textbf{(b)} $\Stomean$ for different $T$.
    Initially there is a slight decrease in insertion loss that could be due to TLS saturation. At higher $T$ transmission breaks down due to quasiparticle losses.
    \textbf{(c)} and \textbf{(d)} $\Sto$ as a function of $f$ for different $T$, both the data and model. Data in the top panel is the same as in the bottom one but plotted on a different vertical scale.
    \textbf{(e)} $\Sto$ as a function of $f$ for a dipstick measurement in liquid helium showing the bandgap frequency. The model including the bandgap is shown as we chose the $\alpha_\text{KI}$ to match the data.
  }
  \label{fig:twpa_loss_vs_temperature}
\end{figure*}

This section presents detailed TWPA insertion loss data as a function of frequency and temperature. 
The insertion loss is similar to that in the setup of Ref.~\cite{Faramarzi2024}. 
\cref{fig:twpa_loss_vs_temperature}(a) shows gradual transmission breakdown, consistent with the niobium ground plane $\Tc$ of approximately \SI{9.1}{\kelvin}.

\cref{fig:twpa_loss_vs_temperature}(b) shows mean insertion loss $\Stomean$ versus temperature and compares to our model (detailed in \cref{sec:twpa_model}). 
The model parameters are primarily based on nominal dimensions and material properties rather than a fit to the data.
Only dielectric loss and a kinetic inductance scaling factor for the NbTiN $\alpha_\text{KI}$ were adjusted to match the data.
The dielectric loss was assumed to be primarily due to the thin dielectric in the inverse microstrip and this was adjusted to match $\Sto(f)$ at low temperature.
$\alpha_\text{KI}$ was adjusted to match bandgap frequency measured in dipstick measurements (\cref{fig:twpa_loss_vs_temperature}(e)).
Dipstick data were acquired before TWPA mounting in the dilution refrigerator.
The need for the $\alpha_\text{KI}$ factor could also be due to slight variations in the NbTiN film thickness or normal state conductivity across a wafer. 
The smoother $\Sto(f)$ in the dipstick data suggests that the impedance mismatches outside the TWPA in the dilution refrigerator (e.g., diplexers and circulators) are non-negligible.
The diplexers prevent the observation of the bandgap in transmission measurements in the dilution refrigerator.
Measuring the bandgap in transmission without diplexers could improve our understanding of temperature (and magnetic field) dependence.

The model matches $\Stomean$ data reasonably well, but we observe notable deviations when considering the full frequency dependence at different temperatures (\cref{fig:twpa_loss_vs_temperature}(c,d)) rather than average insertion loss.
At low temperatures (below \SI{2}{\kelvin}), an initial insertion loss decrease likely reflects two-level fluctuator saturation, which is not included in the model~\cite{Gao2008, Abdisatarov2024}.
At higher temperatures, data show stronger high-frequency losses.

\section{Modeling $\Sto$}
\label{sec:twpa_model}

We model the TWPA transmission $\Stopl$ to understand the temperature and the magnetic field dependence, following Ref.~\cite{Klimovich2024}.
The model uses Mattis-Bardeen theory~\cite{MattisBardeen1958} to calculate superconductor surface impedance and the temperature and field dependence of the superconducting gap.
Whereas the temperature dependence is reasonably described, the magnetic field dependence is more complex and cannot be explained by gap suppression alone.
Field-induced gap suppression without temperature increase does not substantially increase surface resistance. Overall, surface resistance only changes significantly near $\Bc$.
Model limitations in describing magnetic field dependence stem from neglecting vortex effects and screening currents in the ground plane, which increase pair breaking.
More sophisticated models like Ref.~\cite{Kwon2018} could describe the data by combining two-fluid models with a Coffey-Clem model~\cite{CoffeyClem1991} for vortex dynamics and losses.
However, too many free parameters would be required; we do not pursue this approach.

Rather than fitting the model to data, we choose physically reasonable values based on nominal device dimensions and material properties.
We only varied two parameters, the dielectric loss in the thin dielectric layer and a kinetic inductance scaling factor $\alpha_\text{KI}$ for the NbTiN, to match the low-temperature insertion loss and the bandgap frequency measured in a dipstick setup.
Model parameters and values are listed in \cref{tab:model_parameters}.
Model code is provided in the Zenodo repository accompanying this work.
While a full parameter optimization could improve the fit to the data, crucial bandgap frequency data that could constrain the parameters could not be measured due to diplexers in our setup.

The calculation of $\Sto$ for each frequency follows these steps:
\begin{enumerate}
\item Calculate the superconducting gap $\Delta(T,B)$ for both Nb and NbTiN materials (\cref{subsec:temperature_field_dependence_gap}).
\item Compute surface impedances $\Zs = \Rs + j\Xs$ using Mattis-Bardeen theory with the temperature- and field-dependent gaps (\cref{subsec:mb_surface_impedance}).
\item Calculate transmission line parameters (characteristic impedance $\Zzero$, propagation constant $\gamma$) including kinetic inductance contributions (\cref{subsec:microstrip_params}).
\item Use cascaded ABCD matrices to model the complete device with modulated unit cells (\cref{subsec:abcd_matrix}).
\item Convert the total ABCD matrix to the $\Stopl$ parameter.
\end{enumerate}

\subsection{Temperature and field dependence of the gap}
\label{subsec:temperature_field_dependence_gap}

This section explains how we model superconducting gap parameter $\Delta(T,B)$ as a function of temperature and magnetic field, which is the key input parameter for the Mattis-Bardeen surface impedance calculations.
We have estimates for the $\Tc$ of NbTiN and Nb based on our DC characterization and cooldown data (see \cref{sec:dc_characterization_of_material_tc_and_bc} and \cref{sec:tc_based_on_cooldown_data}).
For NbTiN we use $\Tc = \SI{12.0}{\kelvin}$ based on the cooldown data (consistent with \cref{sec:tc_based_on_cooldown_data}), while our DC characterization yielded a slightly higher value, possibly due to variations in film properties across the wafer or different measurement techniques.
For both materials, we assume the BCS gap to critical temperature ratio $\gapratio = 3.5$ as an approximation; the actual values for $\gapratio$ are typically larger and can vary depending on film thickness and quality~\cite{Hong2013,khan2022}.

For the temperature dependence of the gap, we assume~\cite{Gross1986}
\begin{equation*}
  \Delta(T) = \Delta_0 \tanh\left(1.74\sqrt{\frac{\Tc}{T}-1}\right),
\end{equation*}
where $\Tc$ is the critical temperature and $\Deltazero$ is the gap at zero temperature. 
For magnetic field suppression, the functional form depends on field orientation: for fields not too close to the respective critical ones, out-of-plane fields reduce the gap linearly as 
\begin{equation*}
  \Delta(\Bperp) = \Deltazero \left(1 - \frac{\Bperp}{B_{c\perp}}\right),  
\end{equation*}
while in-plane fields cause a Ginzburg-Landau type  gap suppression,
\begin{equation*}
  \Delta(\Bpar) = \Deltazero \sqrt{1 - \left(\frac{\Bpar}{B_{c\parallel}}\right)^2}.
\end{equation*}
where $B_{c\perp}$ and $B_{c\parallel}$ are effective critical magnetic fields in the two directions (the effective critical fields are proportional to, but larger than, the actual ones -- see Ref.~\cite{Janssen2024}).
We note that for the magnetic field dependence, ultimately this form of gap suppression is not sufficient to explain the observed transmission loss, as discussed in \cref{sec:twpa_insertion_loss_vs_magnetic_field}.

\subsection{Mattis-Bardeen Surface Impedance}
\label{subsec:mb_surface_impedance}

We use Mattis-Bardeen theory to calculate the surface impedance $\Zs = \Rs + j\Xs$ for both the NbTiN strip and the Nb ground plane, which provides the electromagnetic losses and kinetic inductance contributions needed for the transmission line parameter calculations.
The key material parameters are the normal state conductivities $\sigman$ and the superconducting gaps $\Delta(T,B)$, as well as the angular frequency $\omega$ and the temperature $T$.
For superconducting materials with gap parameter $\Delta(T,B)$, surface impedance is calculated as:
\begin{equation*}
\Zs = \Rs + j\Xs = \sqrt{\frac{j\omega\mu_0}{\sigma_1 - j\sigma_2}}
\end{equation*}
where $\sigma_1$ and $\sigma_2$ are the real and imaginary parts of the complex conductivity.

The numerical implementation uses dimensionless variables $\tilde{E} = E/\Delta$, $\tilde{\omega} = \hbar\omega/\Delta$, and $\tilde{T} = k_BT/\Delta$, where $E$ is the quasiparticle energy, $\omega$ is the angular frequency, and $T$ is the temperature.
The complete Mattis-Bardeen expressions~\cite{MattisBardeen1958,Tinkham2004} are:
\begin{align*}
  \frac{\sigma_1}{\sigma_n} &= \frac{2}{\tilde{\omega}} \int_1^\infty \frac{[f(\tilde{E}) - f(\tilde{E}+\tilde{\omega})](\tilde{E}(\tilde{E}+\tilde{\omega}) + 1)}{\sqrt{\tilde{E}^2-1}\sqrt{(\tilde{E}+\tilde{\omega})^2-1}} d\tilde{E} \\
  \nonumber & + \frac{\theta(\tilde\omega-2)}{\tilde\omega} \int^{-1}_{1-\tilde{\omega}} \frac{[1 - 2f(\tilde{E}+\tilde{\omega})](\tilde{E}(\tilde{E}+\tilde{\omega}) + 1)}{\sqrt{\tilde{E}^2-1}\sqrt{(\tilde{E}+\tilde{\omega})^2-1}} d\tilde{E}\\
  \frac{\sigma_2}{\sigma_n} &= \frac{1}{\tilde{\omega}} \int_{-1}^{1-\tilde{\omega}} \frac{[1 - 2f(\tilde{E}+\tilde{\omega})](\tilde{E}(\tilde{E}+\tilde{\omega}) + 1)}{\sqrt{1-\tilde{E}^2}\sqrt{(\tilde{E}+\tilde{\omega})^2-1}} d\tilde{E}
\end{align*}
where $f(\tilde{E}) = 1/(e^{\tilde{E}/\tilde{T}}+1)$ is the Fermi-Dirac distribution function, $\sigman$ is the normal state conductivity, and $\theta(\tilde\omega-2)$ is the Heaviside step function.
The second term in $\sigma_1$ represents pair-breaking processes that become active when $\tilde{\omega} > 2$ (i.e. $\hbar\omega > 2\Delta$).
Our implementation neglects this pair-breaking term as it is not significant in our frequency range of interest (\freqrange) where $\tilde{\omega} \ll 1$ for both materials except near $\Tc$ or $\Bc$, so near the transition the model underestimates pair breaking.

The integrals are evaluated numerically using adaptive quadrature with physics-motivated integration points based on the gradient of the integrand and regularization (adding small complex numbers to the integrand) to better handle singularities.
For the thin NbTiN and Nb films used in our device, finite-thickness corrections are applied to account for the reduced dimensionality compared to bulk superconductors, see the next subsection.
The resulting surface impedances $\Zss$ (strip) and $\Zsg$ (ground plane) are input variables for the calculation of the transmission line parameters in the following section.

For magnetic field effects beyond gap suppression, additional terms could be incorporated to account for vortex-induced losses and increased pair breaking, but we found that the experimental data available would not allow for a satisfactory understanding of these microscopic effects.

\subsection{Transmission line parameters of the inverse microstrip}
\label{subsec:microstrip_params}

We calculate the distributed transmission line parameters of the inverse microstrip, characteristic impedance $\Zzero$ and propagation constant $\gamma = \alpha + j\beta$, which are required for the ABCD matrix modeling of the TWPA unit cells.
The inverse microstrip geometry features a superconducting strip positioned between the ground plane and the substrate, in contrast to a conventional microstrip where the strip is on top.
The key geometric parameters of our device are the strip width $w$, dielectric thickness $h$, strip thickness $\ts$, and ground plane thickness $\tg$.
We start with the baseline transmission line parameters calculated using the Hammerstad-Jensen model~\cite{Hammerstad1980}, then apply superconducting corrections to account for the kinetic inductance contributions from both the strip and the ground plane.

The Hammerstad-Jensen model first calculates thickness-corrected widths to account for finite conductor thickness:
\begin{align*}
\Delta u_1 &= \frac{\ts/h}{\pi} \ln\left(1 + \frac{4e}{(\ts/h)\coth^2\left(\sqrt{6.517 u}\right)}\right) \\
\Delta u_r &= \frac{1}{2}\left(1 + \frac{1}{\cosh\sqrt{\epsr - 1}}\right) \Delta u_1
\end{align*}
where $u = w/h$ is the normalized strip width, $\ts$ is the strip thickness, $h$ is the dielectric thickness, and $\epsr$ is the relative permittivity.

The effective dielectric constant for the baseline geometry is:
\begin{align*}
\setlength{\arraycolsep}{10pt} 
&\varepsilon_{\text{eff,0}} = \frac{\epsr + 1}{2} + \frac{\epsr - 1}{2} \left(1 + \frac{10}{u_r}\right)^{-ab} \\
&a = 1 + \frac{1}{49}\ln\left(\frac{u_r^4 + (u_r/52)^2}{u_r^4 + 0.432}\right) + \frac{1}{18.7}\ln\left(1 + \left(\frac{u_r}{18.1}\right)^3\right) \\
&b = 0.564\left(\frac{\epsr - 0.9}{\epsr + 3}\right)^{0.053}
\end{align*}
where $u_r = u + \Delta u_r$ is the thickness-corrected normalized width.
We use separate dielectric properties for the substrate and superstrate: relative permittivities $\epsrsub$ and $\epsrsuper$ and loss tangents $\tandeltasub$ and $\tandeltasuper$.

The effective dielectric constant for mixed dielectrics is calculated by weighting the substrate and superstrate contributions:
\begin{equation*}
\varepsilon_{\text{eff}} = \epsrsub \qsub + \epsrsuper \qsuper
\end{equation*}
where the field filling factors $\qsub$ and $\qsuper$ are defined below.

The characteristic impedance of the baseline microstrip is
\begin{equation*}
Z_{0,\text{base}} = \frac{\etazero}{2\pi\sqrt{\varepsilon_{\text{eff,0}}}} \ln\left(\frac{F_u}{u_r} + \sqrt{1 + \left(\frac{2}{u_r}\right)^2}\right),
\end{equation*}
where $\etazero = 377\,\Omega$ is the free space impedance and 
\begin{equation*}
F_u = 6 + (2\pi - 6)\exp(-(30.666/u_r)^{0.7528}).
\end{equation*}

From the baseline parameters, we extract the geometric inductance and capacitance:
\begin{align*}
L'_{\text{geom}} &= \frac{Z_{0,\text{base}}}{v_{\text{ph,base}}} = \frac{Z_{0,\text{base}}\sqrt{\varepsilon_{\text{eff,0}}}}{c} \\
C' &= \frac{1}{Z_{0,\text{base}} v_{\text{ph,base}}} = \frac{\sqrt{\varepsilon_{\text{eff,0}}}}{Z_{0,\text{base}} c}
\end{align*}

The kinetic inductance contributions are calculated from the surface impedances $\Zss$ and $\Zsg$ and are added to the geometric inductance.
The surface inductance is corrected for finite-thickness films using the Pearl effect:
\begin{equation*}
\mathcal{L}_{\text{surf}} = \mu_0\lambda\coth\left(\frac{t}{\lambda}\right),
\end{equation*}
where $\mu_0$ is the permeability of free space, $\lambda$ is the penetration depth, and $t$ is the film thickness.
The penetration depth $\lambda$ is calculated from the imaginary part of the surface impedance:
\begin{equation*}
\lambda = \frac{\text{Im}(Z_{\mathrm{s},\mathrm{s}})}{\omega \mu_0}, \quad \lambda_{\mathrm{g}} = \frac{\text{Im}(Z_{\mathrm{s},\mathrm{g}})}{\omega \mu_0}
\end{equation*}
where $f$ is the frequency, $\Zss$ is the surface impedance of the strip, and $\Zsg$ is the surface impedance of the ground plane.

The total distributed inductance combines geometric and kinetic components:
\begin{equation*}
L' = L'_{\text{geom}} + \frac{\mathcal{L}_{\text{surf,s}} \alphaKI}{w} + \frac{\mathcal{L}_{\text{surf,g}}}{w}
\end{equation*}
where $\alphaKI$ is a kinetic inductance enhancement factor for the thin NbTiN center conductor.
It was actually one of two parameters we used to vary to match the model to the data. 
Without this factor, the TWPA bandgap in the model would not match the helium dipstick measurement presented in \cref{fig:twpa_loss_vs_temperature} (e). 
This could be due to variations in thickness or conductivity as there are variations in film properties across a wafer.
Additionally, a non-uniform current distribution in the strip could also increase the effective kinetic inductance.
We decided to include this factor rather than to try to adjust the nominal parameters. 
We also assume that the current distribution in the ground plane is uniform with the width of the center conductor.
This is an approximation, and the current will probably spread out, especially once  $\lambda_{\mathrm{g}}$ approaches the ground plane thickness $\tg$.
This would in part offset increases in the kinetic inductance of the ground plane.

We include both dielectric and conductor loss in the real part $\alpha$ of the complex propagation constant $\gamma = \alpha + j\beta$. 
The two loss contributions can be added: $\alpha = \alphad + \alphac$. 
Dielectric losses arise from the finite loss tangents of the substrate and superstrate materials, weighted by their respective field filling factors:
\begin{align*}
\qsub &= \frac{1}{2} + \frac{1}{4}\left(1 - \tanh\left(\frac{1}{2}\ln\left(\frac{\varepsilon_{\text{sub}}}{\varepsilon_{\text{super}}}\right)\right)\right)\frac{1 + \ln(u^2)}{1 + 10/u} \\
q_{\text{super}} &= 1 - q_{\text{sub}} \\
\alphad &= \frac{\pi f}{c}\sqrt{\varepsilon_{\text{eff}}} \left(\qsub\tandeltasub + \qsuper\tandeltasuper\right)
\end{align*}
where $\qsub$ and $\qsuper$ are the field filling factors in the substrate and the superstrate, respectively. 
Conductor losses are calculated from the real parts of the surface impedances:
\begin{equation*}
\alphac = \frac{\Rss/w + \Rsg/w}{\Zzero}\sqrt{\varepsilon_{\text{eff}}}
\end{equation*}
where $\Rss = \text{Re}(\Zss)$ and $\Rsg = \text{Re}(\Zsg)$ are the surface resistances of the strip and the ground plane, respectively.
These losses are incorporated into the complex characteristic impedance by calculating the resistance and conductance per unit length: $R' = \alphac \Zzero$ and $G' = 2\alphad/\Zzero$. The complex characteristic impedance is then calculated:
\begin{equation*} 
Z_0^{\text{complex}} = \sqrt{\frac{R' + j\omega L'}{G' + j\omega C'}}.
\end{equation*} 
These transmission line parameters ($\Zzero$, $\gamma$) serve as inputs for the ABCD matrix formalism described in the next section.

\subsection{ABCD Matrix Formalism to calculate S21}
\label{subsec:abcd_matrix}

Using the transmission line parameters from the previous section (characteristic impedance $\Zzero$ and propagation constant $\gamma = \alpha + j\beta$), we model the TWPA using cascaded ABCD matrices to calculate the overall $\Stopl$.
The TWPA modulation is implemented using a unit cell with periodically spaced capacitive stubs.

For a transmission line segment of length $d$, the ABCD matrix is:
\begin{equation*}
\begin{bmatrix}
A & B \\
C & D
\end{bmatrix}_{\text{line}} = 
\begin{bmatrix}
\cosh(\gamma d) & \Zzero \sinh(\gamma d) \\
\frac{1}{\Zzero}\sinh(\gamma d) & \cosh(\gamma d)
\end{bmatrix}
\end{equation*}
For a capacitive stub of length $\Ls$ with stub admittance $\Ystub = j\frac{2}{\Zzero}\tan(\beta \Ls)$, the ABCD matrix is:
\begin{equation*}
\begin{bmatrix}
A & B \\
C & D
\end{bmatrix}_{\text{stub}} = 
\begin{bmatrix}
1 & 0 \\
Y_{\text{stub}} & 1
\end{bmatrix}
\end{equation*}
The unit cell structure consists of transmission line segments with stub spacing $d$ and contains $n_{\text{stubs}}$ stubs.
The total device contains $n_{\text{cells}}$ unit cells.
The modulation of stub lengths follows a sinusoidal pattern:
\begin{equation*}
L_i = l_0 + l_a \sin\left(2\pi \frac{i}{n_{\text{stubs}}}\right),
\end{equation*}
where $l_0$ is the average stub length and $l_a$ is the modulation amplitude.

The unit cell matrix is constructed by cascading (matrix multiplication) the transmission line segments and stub matrices:
\begin{equation*}
\mathbf{A}_{\text{cell}} = \mathbf{A}_{\text{line,1/2}} \prod_{i=1}^{n_{\text{stubs}}} \left( \mathbf{A}_{\text{stub,i}} \mathbf{A}_{\text{line}} \right)
\end{equation*}
where $\mathbf{A}_{\text{line,1/2}}$ represents the initial half-length transmission line segment.
For the complete TWPA with $n_{\text{cells}}$ unit cells, the total ABCD matrix is computed using matrix exponentiation:
\begin{equation*}
\mathbf{A}_{\text{total}} = \mathbf{A}_{\text{cell}}^{n_{\text{cells}}}
\end{equation*}

Finally, the $\Stopl$ parameter is calculated from the total ABCD matrix assuming a \fiftyohm\, environment:
\begin{equation*}
S_{21} = \frac{2}{A + \nicefrac{B}{\Zenv} + C \Zenv + D}
\end{equation*}
where $\Zenv = \fiftyohm$.

\begin{table}[t]
\centering
\resizebox{\columnwidth}{!}{
\begin{tabular}{lll}
\toprule
Parameter & Value & Description \\
\midrule
\multicolumn{3}{l}{\textbf{TWPA design parameters}} \\
$w$ & \SI{340}{\nano\meter} & Microstrip width \\
$t_{\mathrm{s}}$ & \SI{35}{\nano\meter} & Strip thickness (NbTiN) \\
$t_{\mathrm{g}}$ & \SI{350}{\nano\meter} & Ground plane thickness (Nb) \\
$h$ & \SI{100}{\nano\meter} & Dielectric thickness \\
$d$ & \SI{2.21}{\micro\meter} & Unit cell period \\
$n_{\text{stubs}}$ & 59 & Number of stubs per cell \\
$n_{\text{cells}}$ & 80 & Total number of unit cells \\
$l_0$ & \SI{10.8}{\micro\meter} & Average stub length \\
$l_a$ & \SI{2.08}{\micro\meter} & Modulation amplitude \\
\midrule
\multicolumn{3}{l}{\textbf{Material property parameters}} \\
$\Tc^{\text{NbTiN}}$ & \SI{12.0}{\kelvin} & Critical temperature (NbTiN) \\
$\Tc^{\text{Nb}}$ & \SI{9.15}{\kelvin} & Critical temperature (Nb) \\
$\gapratio$ & 3.5 & Gap ratio (both materials) \\
$\sigmaNbTiN$ & \SI{0.56e6}{\siemens\per\meter} & Normal state conductivity (NbTiN) \\
$\sigmaNb$ & \SI{5e6}{\siemens\per\meter} & Normal state conductivity (Nb) \\
$\alphaKI$ & 1.6 & Kinetic inductance enhancement factor \\
$\epsrsub$ & 10.3 & Substrate relative permittivity \\
$\epsrsuper$ & 11.4 & Superstrate relative permittivity \\
$\tandeltasub$ & \SI{1e-5}{} & Substrate loss tangent \\
$\tandeltasuper$ & 0.03 & Superstrate loss tangent \\
\bottomrule
\end{tabular}
}
\vspace{0.5em}
\caption{Summary of model parameters with chosen values}
\label{tab:model_parameters}
\end{table}

\section{TWPA insertion loss vs magnetic field}
\label{sec:twpa_insertion_loss_vs_magnetic_field}

This section discusses TWPA insertion loss as a function of field for the three orthogonal magnetic field directions. 
Although we did not find a model that describes the data with satisfying accuracy (unlike for the temperature dependence), the main features can be understood qualitatively and generally follow expectations.

\subsection{$\Bparone$ direction}

\begin{figure}
  \centering
  \includegraphics[width=\columnwidth]{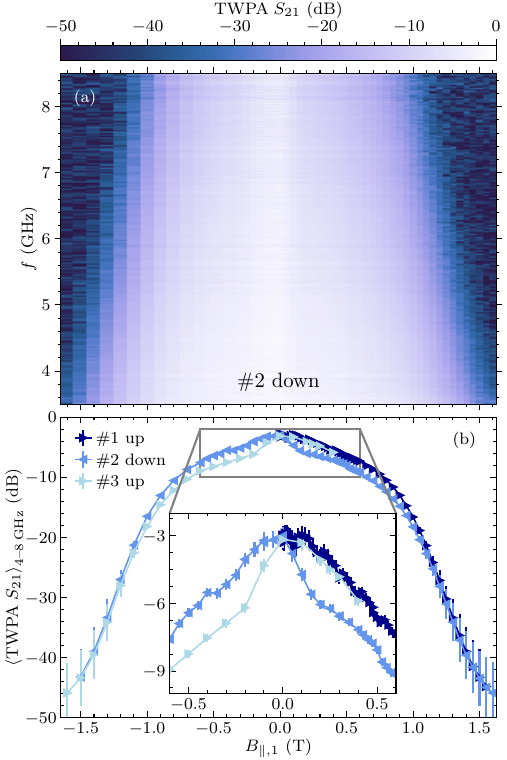}
  \caption{
    \textbf{(a)} $\Sto$ as a function of $\Bparone$ and frequency.
    \textbf{(b)} Average insertion loss $\Stomean$ for different sweeps of $\Bparone$. 
    The TWPA is initialized by a thermal cycle in zero field.
    The field is then ramped to \SI{1.6}{\tesla}, down to \SI{-1.6}{\tesla}, and back up past \SI{0}{\tesla}. 
    The inset shows the magnified low-field region.
  }
  \label{fig:twpa_loss_vs_Bpar1}
\end{figure}

\cref{fig:twpa_loss_vs_Bpar1} shows $\Sto$ for a $\Bparone$ range beyond where the TWPA provides gain (as presented in the main text), up to transmission breakdown.
\cref{fig:twpa_loss_vs_Bpar1}(b) shows mean insertion loss $\Stomean$ for different sweeps, revealing hysteresis that likely indicates that vortices play a role, even at relatively low $\Bparone$.
This at least partially arises due to misalignment of the sample plane to $\Bparone$; this becomes clearer when comparing  it to the $\Bpartwo$ direction presented below.

Mattis-Bardeen modeling with field-dependent gap suppression does not describe the data well, even if we assume that the hysteresis is due to misalignment. 
While $\Stomean$ versus $T$ and $\Bparone$ appear similar, field-dependent models predict sudden changes only near $\Bc$, whereas we observe gradual degradation at much lower fields.
With likely combinations of misalignment-induced vortices, screening currents, and in-plane fields affecting vortex dynamics, we concluded that fitting complex models to field dependence would not provide additional insight.

\subsection{$\Bpartwo$ direction}

\begin{figure}
  \centering
  \includegraphics[width=\columnwidth]{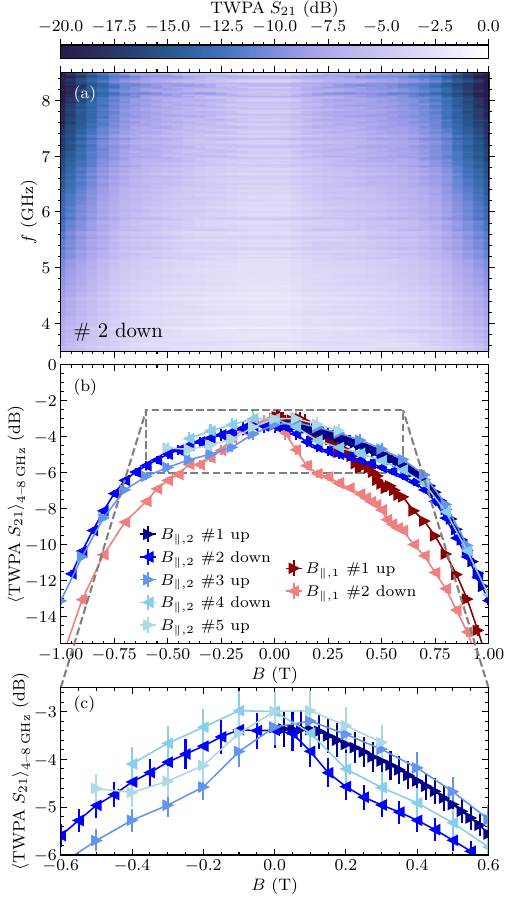}
  \caption{
    \textbf{(a)} $\Sto$ as a function of $\Bpartwo$ and frequency.
    \textbf{(b)} $\Stomean$ as a function of $\Bpartwo$ and $\Bparone$ for comparison.
    Interestingly, the second in-plane direction shows less hysteresis and a slower suppression of $\Stomean$. 
    The different behavior in the two in-plane directions could be due to alignment or due to device geometry playing a role that we do not understand yet.
    \textbf{(c)} $\Stomean$ as a function of $\Bpartwo$ in the low-field region. 
    When the field is swept back and forth with smaller and smaller amplitudes toward zero $\Stomean$ can be slightly improved.
    This suggests that with thermal cycles at nominally zero field, the device likely has some vortex losses.
  }
  \label{fig:twpa_loss_vs_Bpar2}
\end{figure}

For the second in-plane direction, we acquired only transmission data (no gain or $\DSNR$ measurements).
Data are shown in \cref{fig:twpa_loss_vs_Bpar2}.
Vector magnet limitations restricted the $\Bpartwo$ field range.
\cref{fig:twpa_loss_vs_Bpar2}(b) shows mean $\Stomean$ versus both $\Bpartwo$ and $\Bparone$.
Field dependence is similar, but $\Stomean$ suppression is less pronounced and hysteresis smaller for $\Bpartwo$, possibly indicating better in-plane alignment for that direction.
However, reduced hysteresis could also be a result of the smaller maximum field before reversing scan direction.
Given the TWPA transmission line meandering, neither $\Bparone$ nor $\Bpartwo$ direction is perfectly aligned with device current flow, and differences between the field directions are not expected except due to a difference in misalignment.

\subsection{$\Bperp$ direction}

\begin{figure}
  \centering
  \includegraphics[width=\columnwidth]{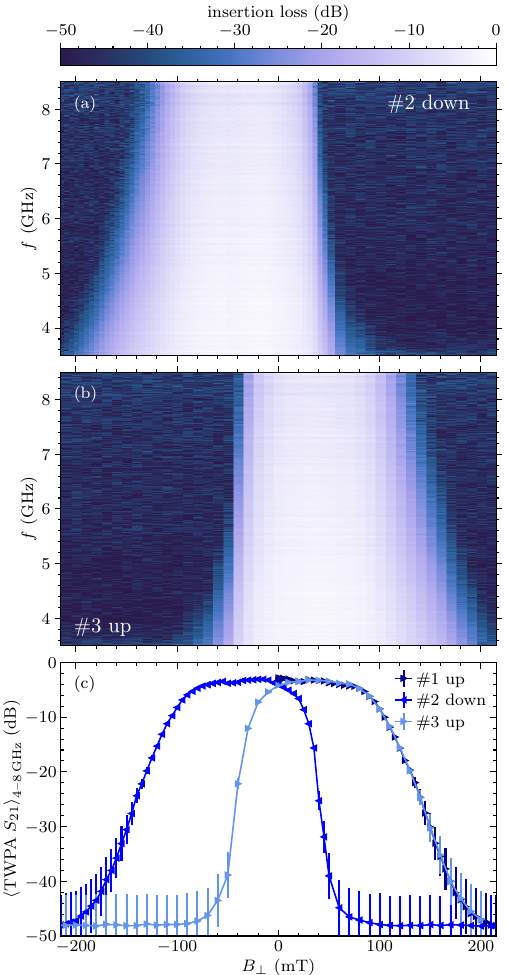}
  \caption{
    \textbf{(a)} and \textbf{(b)} $\Sto$ as a function of $\Bperp$ and frequency for a down and subsequent up sweep.
    Hysteresis is clearly strongest for $\Bperp$, as expected.
    \textbf{(c)} $\Stomean$ for different sweeps of $\Bperp$. 
  }
  \label{fig:twpa_loss_vs_Bperp}
\end{figure}

The $\Bperp$ dependence of the TWPA insertion loss is shown in \cref{fig:twpa_loss_vs_Bperp}.
\cref{fig:twpa_loss_vs_Bperp} (b) shows the strong hysteresis of $\Stomean$ as a function of $\Bperp$.
This hysteresis is similar in resonators~\cite{Bothner2012}.
TWPA transmission is fully suppressed around \SI{200}{\milli\tesla} and only starts to recover around \SI{80}{\milli\tesla} in the downward sweep, and peaks beyond zero field.
One can recover the initial transmission by thermal cycling, which we did between all separate field sweeps. One could also sweep back and forth with smaller and smaller amplitudes to recover the initial transmission~\cite{Bothner2012} as we saw for the $\Bpartwo$ direction in \cref{fig:twpa_loss_vs_Bpar2} (c).

\section{DC characterization of material $\Tc$ and $\Bc$}
\label{sec:dc_characterization_of_material_tc_and_bc}

\begin{figure}
  \centering
  \includegraphics[width=\columnwidth]{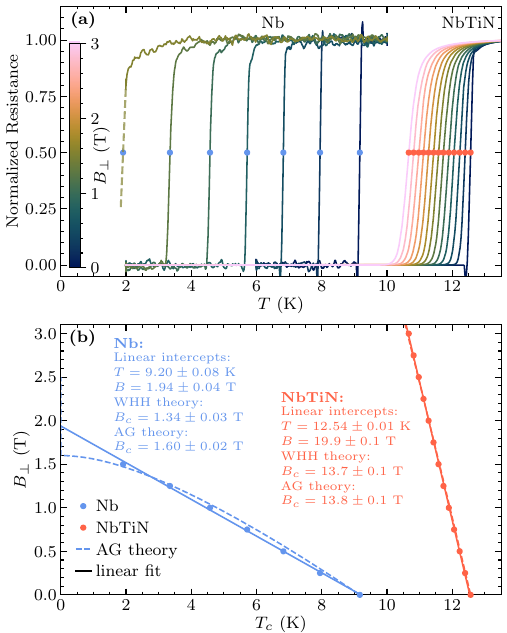}
  \caption{
    \textbf{(a)} Normalized resistance as a function of temperature $T$ for Nb and NbTiN films from a similar device at different $\Bperp$.
    $\Tc$ is extracted from the midpoint of the transitions.
    For $\Bperp=\SI{1.5}{\tesla}$, the highest field datapoint for Nb, the transition could not be fully observed as the minimum temperature of the measurement system was \SI{2}{\kelvin} and the data was extrapolated (dashed line) to get an estimate.
    \textbf{(b)} $\Tc$ at different $B$ shows a linear decrease with increasing field, consistent with a disordered superconductor.
    We can estimate $\Bc$ and $\Tc$ based on the axis intersects of a linear fit for the two superconductors or the AG model described in the text; for NbTiN the two fits are indistinguishable in the range of the plot. 
  }
  \label{fig:dc_measurements_Tc_Bc}
\end{figure}

To determine $\Tc$ and $\Bc$ of the materials used in the TWPA, nominally identical Nb and NbTiN films were deposited and measured in a Physical Property Measurement System (PPMS). 
Four-terminal resistance measurements were acquired as functions of temperature $T$ and magnetic field $\Bperp$.
We extract $\Tc$ from the superconducting transition midpoint (\cref{fig:dc_measurements_Tc_Bc}(a)).
$\Tc$ versus $\Bperp$ (\cref{fig:dc_measurements_Tc_Bc}(b)) shows a linear decrease with increasing field, similar to other thin film superconductors~\cite{Semenov2009,Zhang2016}. 
We estimate the zero-temperature critical field in two ways. First, using proper extensions of the Abrikosov-Gorkov (AG) theory of $\Tc$ suppression by magnetic impurities (see Ch.10 in Ref.~\cite{Tinkham2004}), or equivalently the results of Werthamer-Helfand-Hohenberg (WHH)~\cite{Werthamer1966} for weakly coupled BCS superconductors in the dirty limit, the zero-temperature upper critical field can be related to its temperature derivative near $\Tc$:
\begin{equation*}
B_{\text{c2}}(0) = 0.69\Tc \left[ \frac{dB_{\text{c2}}}{dT}\right]_{T=\Tc}.
\end{equation*}
From a linear fit to the data for Nb we thus obtain a critical field of \SI{1.34}{\tesla}; this is slightly smaller than the field $\sim$\SI{1.5}{\tesla} at which we observe full suppression of $\Sto$. This may be due to our assumption of negligible spin-paramagnetic effects and spin-orbit interactions. Moreover, the 0.69 prefactor may not be accurate for niobium films~\cite{Zaytseva2020}.

As an alternative to the linear fit, we fit the data using the AG relationship between $\Tc$ and the parameter $\alpha$ characterizing the strength of the relevant pair breaking mechanism~\cite{Tinkham2004}, using
\begin{equation*}
\ln\left(\frac{T_c}{T_{c0}}\right) = \psi\left(\frac{1}{2}\right) - \psi\left(\frac{1}{2} + \frac{\alpha}{T_c/T_{c0}}\right),
\end{equation*}
where $\psi(x)$ is the digamma function and $T_{c0} = T_c(0)$ is the critical temperature in the absence of pair breaking, $\alpha=0$.
For thin films in perpendicular fields, $\alpha = De\Bperp/c$ where $D$ is the electronic diffusion constant and $\Bperp$ is the magnetic field~\cite{Tinkham2004}.
This makes $\alpha$ linearly proportional to the applied field, with the proportionality constant determined by  material properties. In practice, we write $\alpha$ in the form $\alpha=\Bperp/\Bc$ and implement this model numerically, solving the transcendental equation and fitting our experimental data with $B_c$ as the only free parameter while keeping $T_c(0)$ fixed at the measured zero-field value.
This approach gives $\Bc\simeq\SI{1.6}{\tesla}$, a value in better agreement with the observed suppression of $\Sto$, even though the model does not fit the data as well as the simpler linear fit.
Measurements also indicate that NbTiN critical field is large enough that it will not limit TWPA performance, since the fitted AG value of $\Bc$ for NbTiN is approximately 13.8~T (close to the linear fit intersect multiplied by 0.69). 
From this critical field value, we estimate the coherence length $\xi$ via the relation $B_c = \phi_0/2\pi \xi^2$, finding $\xi \simeq 4.9\,$nm.

\section{$\Tc$ estimate based on cooldown data}
\label{sec:tc_based_on_cooldown_data}

\begin{figure}
  \centering
  \includegraphics[width=\columnwidth]{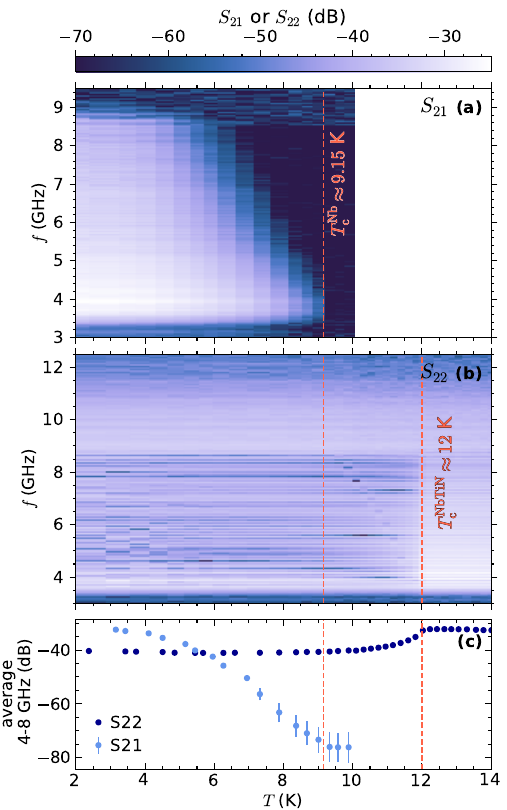}
  \caption{
    Transmission $S_{21}$ through \textbf{(a)} and reflection $S_{22}$ on \textbf{(b)}  the TWPA.
    The critical temperatures of the different materials are indicated.
    Transmission breaks down as the Nb ground plane transitions to its non-superconducting state.
    The reflection data clearly shows when the NbTiN microstrip transitions.
    }
  \label{fig:cooldown_S21_S22}
\end{figure}

We can also estimate critical temperatures of NbTiN and Nb films from $\Stopl$ and $S_{22}$ measurements during TWPA cooldown.
$S_{22}$ is measured using the circulator at the TWPA output (\cref{fig:wiring_diagram}).
Temperature dependence of $\Sto$ (\cref{fig:cooldown_S21_S22}(a)) shows transmission only appears below \SI{9.15}{\kelvin}, consistent with a Nb ground plane $\Tc$ (see \cref{sec:dc_characterization_of_material_tc_and_bc}).

Reflection $S_{22}$ (\cref{fig:cooldown_S21_S22}(b)) shows a clear transition around \SI{12.0}{\kelvin}, close to the NbTiN microstrip $\Tc$ but slightly lower than the DC characterization data on nominally identical films.
Full reflection before NbTiN superconducting transition is consistent with the resistive conducting microstrip representing extreme impedance mismatch.
However, low-temperature frequency dependence sudden change in $S_{22}$ around \SI{8.5}{\giga\hertz} results from the diplexer and has nothing to do with the TWPA.
Curiously, temperature-dependent photonic bandgap frequency likely appears as a faint reflection peak between \SI{9}{\kelvin} and \SI{12}{\kelvin}.

\section{Field and temperature dependence of setup transmission without the TWPA}
\label{sec:field_and_temperature_dependence_of_transmission_without_twpa}

The question arises whether setup components have magnetic field or temperature dependence that could be conflated with the TWPA effects.
Circulators are natural candidates, as is the superconducting NbTi cable between the mixing chamber and the \SI{4}{\kelvin} stage.
Circulators have magnetic shields, but these primarily protect other components from circulator stray fields rather than protecting the circulator from external fields; the shields fail at relatively low fields.
HEMT amplifiers could also show magnetic field dependence, but they are mounted far from the magnet center, at the \SI{4}{\kelvin} stage.
Stray fields in our setup were previously reported in Ref.~\cite{Janssen2024}.

\begin{figure}
  \centering
  \includegraphics[width=\columnwidth]{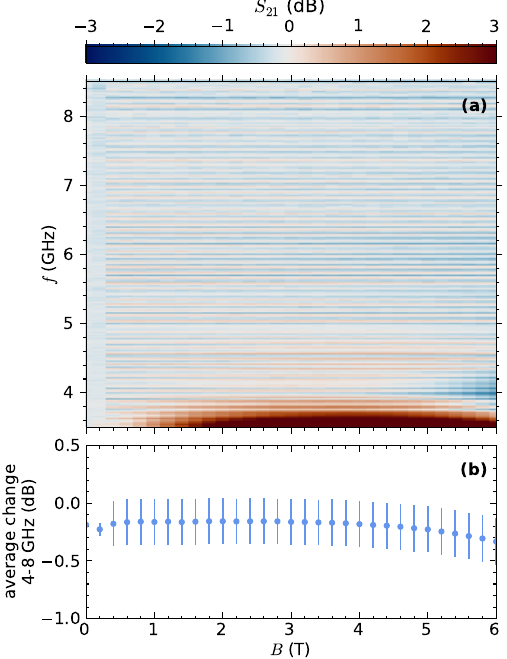}
  \caption{
    \textbf{(a)} Change in $\Stopl$ for a bypass cable instead of the TWPA showing that magnetic field $\Bparone$ has little effect on the calibration.
    At low frequencies, the circulator band edge seems to shift down at higher field actually improving transmission.
    \textbf{(b)} $\Stoplmean$ as a function of $T$ compared to the calibration dataset.
    There is a loss of a fraction of a dB, but this might have been a slightly loose cable.
    At the highest fields some losses are increasing.
  }
  \label{fig:bypass_vs_Bz}
\end{figure}

To estimate these effects, we also measured the bypass configuration, where the TWPA was replaced by a short cable, as a function of magnetic field and temperature.
The field dependence of bypass configuration $\Stopl$ is shown in \cref{fig:bypass_vs_Bz}.
We chose the $z$ direction as it corresponds to the \SI{6}{\tesla} field direction of our vector magnet.
Measurements differ by a fraction of a dB from the calibration measurements, which were taken on a different occasion.
However, except for frequencies below \SI{4}{\giga\hertz} (circulator band edge), almost no field dependence is visible.
Minimal hysteresis can be observed (we also measured when sweeping the field back down), but overall effects appear negligible, particularly since we exclude data below \SI{4}{\giga\hertz} for TWPA figures of merit.

\begin{figure}
  \centering
  \includegraphics[width=\columnwidth]{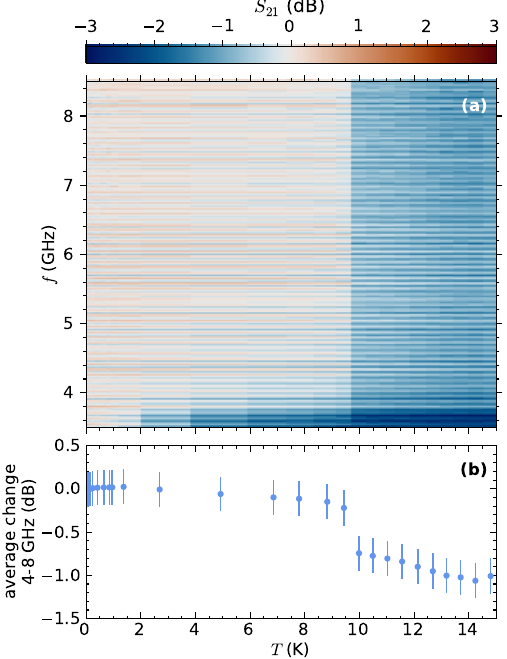}
  \caption{
    \textbf{(a)} Change in $\Stopl$ for a bypass cable instead of the TWPA, showing that temperature $T$ will not affect the calibration in the relevant range.
    \textbf{(b)} $\Stoplmean$ as a function of $T$ compared to the calibration dataset.
    Around \SI{10}{\kelvin} there is a sudden jump likely due to the transition of the superconducting coaxial cable between the mixing chamber and the HEMT amplifier at the \SI{4}{\kelvin} stage.
    There is also likely a slight shift in the circulator band edge with $T$.
  }
  \label{fig:bypass_vs_T}
\end{figure}

The temperature dependence of bypass configuration $\Stopl$ (\cref{fig:bypass_vs_T}) shows weak effects that similarly do not influence TWPA characterization.
Small circulator band edge changes occur around \SI{2}{\kelvin} but only for frequencies below \SI{4}{\giga\hertz}.
A jump at \SI{10}{\kelvin} is consistent with the superconducting transition of the NbTi cable between the mixing chamber and the HEMT amplifier at the \SI{4}{\kelvin} stage
At this temperature, the TWPA shows no transmission, so this does not affect TWPA characterization.

\clearpage
\bibliographystyle{apsrev4-2}
\bibliography{references}

\end{document}